\documentstyle{europhys1}
\input psfig
\begin{document}

\euro{xx}{x}{x-x}{xxxx}
\Date{x xxxxxx xxxx}

\shorttitle{Y.~ASHKENAZY {\it et al.}: SCALE SPECIFIC AND SCALE
INDEPENDENT MEASURES OF HEART RATE}

\title{Scale Specific and Scale Independent Measures of Heart Rate 
Variability as Risk Indicators}

\author{Y.~Ashkenazy\inst{1,2}\footnote{Email:ashkenaz@argento.bu.edu},
M.~Lewkowicz\inst{3,1}, J.~Levitan\inst{3,1,4}, S.~Havlin\inst{1,2},
K.~Saermark\inst{4}, H.~Moelgaard\inst{5}, P.E.~Bloch~Thomsen\inst{6},
M.~Moller\inst{7}, U.~Hintze\inst{7}, and H.V.~Huikuri\inst{8}} 

\institute{
	\inst{1} Dept. of Physics, Bar-Ilan University, Ramat-Gan, Israel\\ 
	\inst{2} Gonda Goldschmied Center, Bar-Ilan University,
	Ramat-Gan, Israel\\  
	\inst{3} Dept. of Physics, College of Judea and Samaria, Ariel,
	Israel\\  
	\inst{4} Dept. of Physics, The Technical University of Denmark,
	Lyngby, Denmark\\ 
	\inst{5} Dept. of Cardiology, Skejby Sygehus, Aarhus University
	Hospital, Aarhus, Denmark\\ 
	\inst{6} Dept of Cardiology, Amtssygehuset i Gentofte,
	Copenhagen University Hospital, Denmark\\ 
	\inst{7} Inst. Clin. Res. Cardiology, University of Southern
	Denmark, Odense, Denmark\\
	\inst{8} Div. Cardiology, Dept. of Medicine, University of Oulu,
	Finland
}

\rec{}{}

\pacs{
\Pacs{05}{45.Tp}{Time series analysis}
\Pacs{05}{40.-a}{Fluctuation phenomena, random processes, noise, and
Brownian motion} 
\Pacs{87}{.}{Biological and medical physics}
\Pacs{87}{19.Hh}{Cardiac dynamics}
}

\maketitle

\begin{abstract}
{
We study the Heart Rate Variability (HRV) using scale specific variance
and scaling exponents as measures of healthy and cardiac impaired
individuals. Our results show that the variance and the scaling exponent
are uncorrelated. We find that the variance measure at certain scales is
well suited to separate healthy subjects from heart patients. However,
for cumulative survival probability the scaling exponents outperform the
variance measure. Our risk study is based on a database containing
recordings from 428 individuals after myocardial infarct (MI) and on
database containing 105 healthy subjects and 11 heart patients. The
results have been obtained by applying three recently developed methods
(DFA - Detrended Fluctuation Analysis, WAV - Multiresolution Wavelet
Analysis, and DTS - Detrended Time Series analysis) which are shown to
be highly correlated. 
}
\end{abstract}

The study of heart rate variability (HRV) has been in use for the last
two decades as part of clinical and prognostic work; international
guidelines for evaluating conventional HRV-parameters do exist
\cite{Task}. The conventional parameters are power spectra
\cite{Akselrod} and standard deviation \cite{Wolf,Ivanov98}. Recently three new
methods of analyzing heart interbeat interval (RR) time series have been
developed, all of them showing signs of improved diagnostic
performance. The three methods are: Detrended Fluctuation Analysis (DFA)
\cite{Peng95,Ivanov99,Pikku,Maki,Bunde00}, Multiresolution Wavelet
Analysis (WAV) 
\cite{Arneodo,Ivanov96,Akay,Thurner98a,Amaral98,Roach} and Detrended
Time Series Analysis (DTS) \cite{Ashkenazy99}. The question which method
and which measure yield better separation between cardiac impaired and
healthy subjects has been recently debated
\cite{Thurner98a,Amaral98,Thurner98b}.

In this Letter we show that variance, which is a scale specific measure,
is well suited to separate between healthy subjects and heart patients.
However, for the myocardial infarct (MI) group the scaling exponent,
which is a scale independent measure, serves as a better risk
indicator. Moreover, we show that the three above mentioned methods for
both variance and scaling exponent, are correlated and converge to
similar results while the variance and the scaling exponent are
uncorrelated.

In our study we use two groups, the MI group, containing 428 heart
patients after MI and a control group, consisting of 105 healthy
individuals and  11 cardiac impaired patients (9 diabetic patients, one
diabetic patient after myocardial infarct, and one heart transplanted
patient). Our analysis is based on 24 hour heart interbeat interval time
series \cite{remark0}. We applied the following methods. 

{\it The DFA Method.} The detrended fluctuation analysis was proposed by
Peng {\it et al} \cite{Peng95}. This method avoids spurious detection of
correlations that are artifacts of nonstationarity. The interbeat
interval time series is integrated after subtracting the global average
and then divided into windows of equal length, $n$. In each window the
data are fitted with a least square straight line which represents the
local trend in that window. The integrated time series is detrended by
subtracting the local trend in each window. The root mean square
fluctuation, the standard deviation $\sigma_{\rm dfa}(n)$ of the
integrated and detrended time series is calculated for different scales
(window sizes); the standard deviation can be characterized by a scaling
exponent $\alpha_{\rm dfa}$, defined as $\sigma(n) \sim n^\alpha$. 

{\it The WAV Method.} In the WAV method
\cite{Thurner98a,Amaral98,Ashkenazy98} one finds the wavelet
coefficients $W_{m,j}$, where $m$ is a `scale parameter' and $j$ is a
`position' parameter (the scale $m$ is related to the number of data
points in the window by $n=2^m$ \cite{remark0}), by means of a wavelet
transform. The standard deviation $\sigma_{\rm wav}(m)$ of the wavelet
coefficients $W_{m,j}$ across the parameter $j$ is used as a parameter
to separate healthy from sick subjects. The corresponding scaling
exponents is denoted by $\alpha_{\rm wav}$.

{\it The DTS Method.} The detrended time series method was suggested in
\cite{Ashkenazy99}. In this method one detrends the RR time series by
subtracting the local average in a running window from the original time
series, resulting in a locally detrended time series. The standard
deviation $\sigma_{\rm dts}$ is calculated for various window scales
with a scaling exponent $\alpha_{\rm dts}$. 

\begin{figure}
\centerline{\psfig{figure=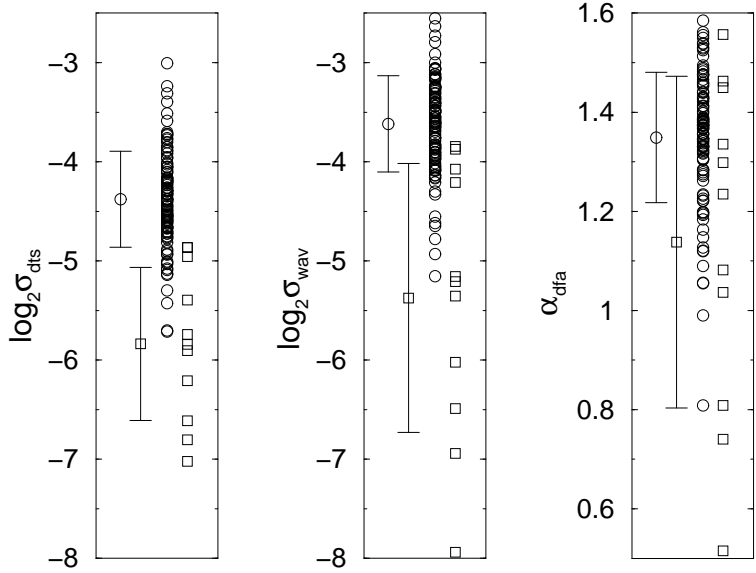,height=3.truein,angle=-90}}
\caption{ 
A comparison between different HRV methods (DTS, WAV, 
and DFA). The 105 healthy subjects are denoted by $\circ$, while 11
heart patients are denoted by $\Box$. 
The error bars indicate the average $\pm$ 1 standard  deviation of each
group. The $\sigma_{\rm dts}$ is calculated at scale $m=8$, $\sigma_{\rm
wav}$ at scale $m=4$, and $\alpha_{\rm dfa}$ for scales $1 \le m \le 4$;
it was previously reported in the literature that this choice of
parameters yields the best separation between healthy group and heart
failure group \protect\cite{Peng95,Thurner98a,Ashkenazy99}.
}
\label{fig1}
\end{figure}

The first suggestion to use a scale independent measure of the HRV as a 
separation parameter was by Peng {\it et al} \cite{Peng95} who found
that a critical value of the DFA scaling exponent $\alpha_{\rm dfa}$ can
distinguish between healthy individuals and heart patients. Thurner {\it
et al} \cite{Thurner98a} used the scale specific WAV variance
$\sigma_{\rm wav}$ in order to better separate the same two groups. 
The debate, which method performs better was continued in two recent
Letters \cite{Amaral98,Thurner98b}. Later on, another independent study
on different database \cite{Ashkenazy98} yielded a better separation
using the scale specific $\sigma_{\rm wav}$ measure.  

In Fig. 1 we compare the conventional measures for HRV for the control group:
the variance (which is calculated for a fixed scale) for the DTS and WAV
method ($\sigma_{\rm dts}$ and $\sigma_{\rm wav}$) and the scaling
exponent (which is calculated for a range of scales) for the DFA method
($\alpha_{\rm dfa}$). One notes that the scale specific measures,
$\sigma_{\rm dts}$ and $\sigma_{\rm wav}$, yield a nearly perfect
separation between healthy and sick subjects (the $p$ value of the
student t-test is less than $10^{-14}$), compared with the scale independent
measure $\alpha_{\rm dfa}$ which yields less pronounced separation ($p$
value of the student t-test is less than $10^{-4}$).  

\begin{figure}
\centerline{\psfig{figure=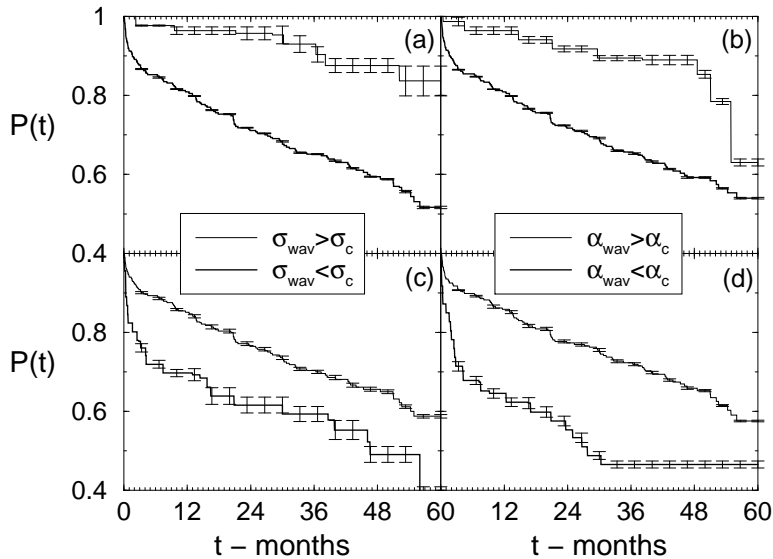,height=3.truein,angle=-90}}
\caption{ 
Cumulative survival probability curves using the WAV
method. We divide the entire group of 328 individuals into two groups
according to a critical value $\sigma_{\rm c}$ or $\alpha_{\rm c}$.
The survival curves shown in the figure are the average of 10 different
survival curves, with close critical values. In the upper
panel we average over the 40-50 largest parameter values and in the
lower panel we average over the 40-50 smallest parameter values. We
perform this average procedure in order to check the sensitivity to the
critical value. The error bars indicate the standard deviation from the
average. The average critical values are (a) $\langle
\log_2\sigma_{\rm c} \rangle = -3.749 $, (b) $\langle \alpha_{\rm c}
\rangle = 0.68$, (c) $\langle \log_2\sigma_{\rm c} \rangle = -5.485 $,
and (d) $\langle \alpha_{\rm c}\rangle = 0.135$.
}
\label{fig2}
\end{figure}

This outcome is reversed when we applied the measures on the MI
group. Since we have no diagnostics on this group, but rather do know
the follow-up history for 328 individuals from the total 428 individuals
of the larger group, we investigate the survival probability of these
subjects  as expressed through the so-called survival curve
\cite{remarkA}. In these 
curves one divides the entire group by means of a specific value of the
$\sigma$ or $\alpha$ measure, called the critical value $\sigma_c$ or
$\alpha_c$. For each subgroup we calculate the cumulative survival
probability given by $P(t+\Delta t)=P(t)[1-\Delta N /N(t)]$, where
$P(t)$ is the probability to survive up to $t$ days after the ECG
recording, $N(t)$ denotes the number of individuals alive at $t$ days
after the examination, and $\Delta N$ is the number of individuals who
died during the time interval $\Delta t$. In Fig. \ref{fig2} we show a
comparison of survival curves where the separating measure in figures
(a) and (c) is the critical standard deviation $\sigma_c$ and in figures
(b) and (d) the critical scaling exponent $\alpha_c$. Individuals with
$\sigma > \sigma_c$ (or $\alpha > \alpha_c$) belong to the subgroup with
the higher survival probability; the upper panel extracts the subgroup
with a high survival probability, whereas the lower panel extracts the
subgroup with a low survival probability. This comparison shows
that the scale independent scaling exponent $\alpha$ serves as a better
prognostic predictor than the scale specific variance $\sigma$ (although
Fig. \ref{fig2}a and b are similar the survival curves of
Fig. \ref{fig2}d are more separated than the survival curves of
Fig. \ref{fig2}c). 

In Fig. \ref{fig2} we use the $\sigma$ and $\alpha$ measures obtained
through the WAV method. However, as we show below all three methods
discussed above are highly correlated and no significant difference is
noticeable in the survival curves when using DFA and DTS measures. 

\begin{figure}
\centerline{\psfig{figure=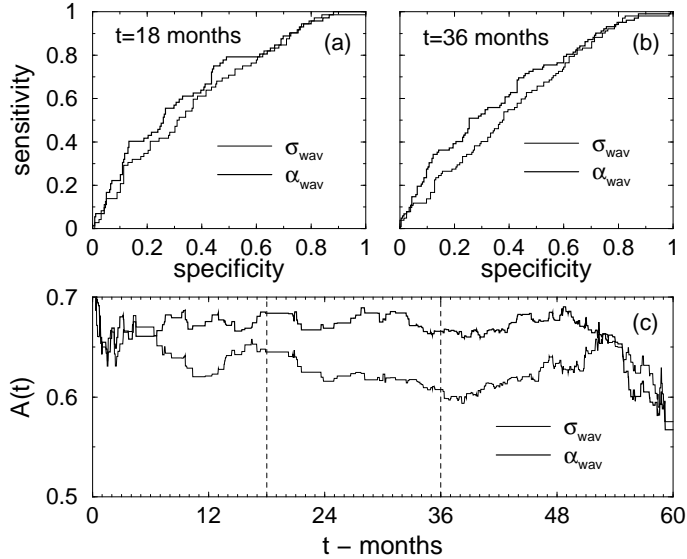,height=3.truein,angle=-90}}
\caption{
The ROC curves (sensitivity vs specificity) of the scale specific,
$\sigma_{\rm wav}$, and scale independent, $\alpha_{\rm wav}$, measures. 
(a) $t=18$ months, (b) $t=36$ months, and (c) $A(t)$ --- the area under
the ROC curve --- as a function of time. In all figures the
scale independent curve is located above the scale specific curve. Thus,
the scale-independent measure is more suitable for prognostic purposes.
}
\label{fig3}
\end{figure}

The advantage of the scale independent measure $\alpha$ over the scale
specific measure $\sigma$ is also shown in  
Fig. \ref{fig3}. Here we scan the possible critical values by the
Receiver Operating Characteristic (ROC) analysis \cite{Swets88}; this
analysis is usually used as a medical diagnostic test and also was the
basic diagnostic test of Ref. \cite{Thurner98b}. The idea of the ROC
method is to compare the result of medical test (positive or negative)
with the clinical status of the patient (with or without disease). The
efficiency of the medical test is judged on the basis of its sensitivity
(the proportion of diseased patients correctly identified) and its
specificity (the proportion of healthy patients correctly
identified). The ROC curve is a graphical presentation of sensitivity
versus specificity as a critical parameter is swept. In our case the
patient status is determined according to its mortality (death or
survival up to time $t$) and according to its mortality
prediction (a patient with parameter value smallest than the critical
value is predicted to die while a patient with a parameter value larger
than the critical value is predicted to survive). In Fig. \ref{fig3}a
and b we present two examples of the ROC 
curves in different times (18 months and 36 months). In both cases the ROC
of the scale independent ($\alpha_{\rm wav}$) curve is located above the scale
specific ($\sigma_{\rm wav}$) curve; the larger the area under the
ROC curve is, the better is the parameter. In the ideal case a patient
with small parameter value will die before the patient with higher
parameter value. In this case the area under the ROC curve will be 1. On
the other hand, when there is no relation between the value of the
parameter and the mortality of the patient the area under the ROC curve
will be 1/2. In Fig. \ref{fig3}c we show the area under the ROC curves
as a function of time ($A(t)$). The scale independent 
($\alpha_{\rm wav}$) curve is located above the scale specific
($\sigma_{\rm wav}$) curve. Thus, the scale independent measure $\alpha_{\rm
wav}$ is more suitable for prognosis. 


\begin{figure}
\centerline{\psfig{figure=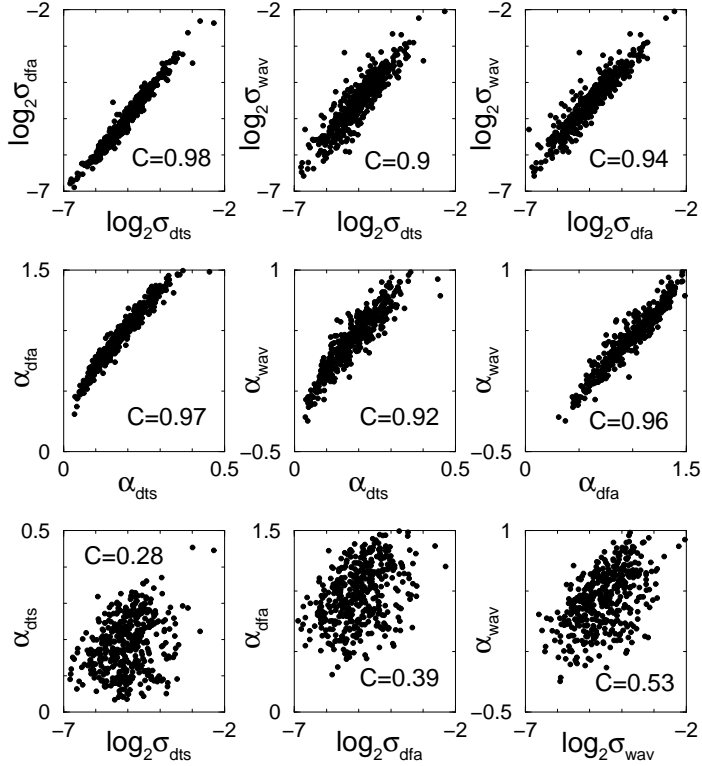,height=4.truein,angle=0}}
\caption{
A comparison between different HRV methods using 
428 individuals. 
The value $C$ in each figure indicates the cross-correlation value
between the different measures \protect\cite{remark3}. 
Upper panel - $\sigma$ measure versus $\sigma$ measure;
middle panel - $\alpha$ measure versus $\alpha$ measure; lower panel -
$\sigma$ measure versus $\alpha$ measure. The $\sigma_{\rm dfa}$ and 
$\sigma_{\rm wav}$ is calculated at $m=4$, $\sigma_{\rm dts}$ at $m=8$;
$\alpha_{\rm dfa}$ and $\alpha_{\rm wav}$ is calculated for $m=1$ to 4, 
$\alpha_{\rm dts}$ for $m=1$ to 8.
}
\label{fig4}
\end{figure}

In order to investigate if the three methods we use are correlated, we apply 
them on the larger MI group consisting of 428 subjects. The top panel of
Fig. \ref{fig4} shows that the variances (the scale specific measure)
of the three methods are highly correlated. This is also true for the scaling 
exponents (the scale independent measure, middle panel). These
comparisons indicate that indeed the various 
methods yield the same results in terms of variance and scaling exponents.
On the other hand, the lower panel of Fig. \ref{fig4} shows that the scale 
specific variance and the scale independent scaling exponent are 
uncorrelated for the DTS and DFA methods and are only faintly correlated
for the WAV method.

From this we conclude
that the $\alpha$ and $\sigma$ measures characterize the interbeat interval
series in different ways; the variance, which is a measure in the time domain 
(and thus is almost invariant to shuffling \cite{Thurner98a}), performs
better as a diagnostic tool, while the
scaling exponent, which is a measure in the frequency domain, depends
on the order of events and performs better as a prognostic tool.
Thus we suggest that the scale specific variance reflects changes in
either the sympathetic or the parasympathetic activities of the
neuro-autonomic nervous system \cite{remark4} which affect the cardiac
ability of 
contraction; the scale specific variance may hint on the instant
condition of the physical properties of the heart. From the above we
also suggest that the scale independent scaling exponent 
characterize the memory interplay of the two competing branches of the
autonomic nervous system (the sympathetic and the parasympathetic
systems) and is thus an expression of the underlying mechanism of heart
regulation (which influences the conventional power spectrum
\cite{Akselrod})\cite{remark5}.   

\vskip12pt
\centerline{***}

We wish to thank Nachemsohns Foundation for financial support.


\begin{thebibliography}{99}
\bibitem{remark0} Note that our wavelet analysis is not restricted to
series lengths which are power of two; in the present study the maximum
scale is $m=4$ ($n=2^4=16$) and thus the series length should be a
multiple of $16$.
\bibitem{remarkA} We also performed a more simple test to verify
whether the scale independent measure is better than the scale specific
measure for prognostication. We divide the group of 328 subjects
into two subgroups --- a group of patients which survive and a group of
patients which died. Then we used the student t-test to compare between
the two groups. For the scale specific $\sigma_{\rm wav}$ we obtained
(average $\pm$ 1 standard deviation) for the alive group $-4.47\pm
0.75$ and the death group $-4.79\pm 0.74$ ($p<0.0003$). For the scale
independent $\alpha_{\rm wav}$ we obtained for the alive group $0.46\pm
0.24$ and for the death group $0.34\pm 0.23$ ($p<8\times
10^{-6}$). Thus, the scale independent measure performs better
prognostics.
\bibitem{remark3} Given two series ($\{x_i\}$ and $\{y_i\}$) with the same
length the cross-correlation value is $C=\sum [(x_i-\langle x
\rangle )(y_i-\langle y \rangle )]/\sqrt{\sum (x_i-\langle x \rangle )^2
\sum (y_i-\langle y \rangle )^2} $, where $\langle \dots \rangle$ indicates
the average of the series.
\bibitem{remark4} The sympathetic/parasympathetic system is the system
which increases/decreases the heart rate.
\bibitem{remark5} An increase/decrease in one of the activities is usually
compensated by a decrease/increase in the other activity. In cardiac
failure the regulation between the two activities breaks down.
\end{thebibliography}
\end{document}